# Development of four kinds of waveguide power divider for S band [*]

HE Xiang(贺祥)[1], HOU Mi(侯泊)[1], ZHANG Jing-Ru(张敬如)[1], CHI Yun-Long(池云龙)[1]

[1] Key Laboratory of Particle Acceleration Physics and Technology, Institute of High Energy Physics,

Chinese Academy of Sciences, Beijing 100049, China

**Abstract:** Four kinds of waveguide power dividers with different structures working at 2.856 GHz are developed. By comparing the simulation performances of these four structures, the power divider with matching rod in the middle of its structure has got the best performance and finally be chosen for fabrication. The prototype has got qualified microwave measurement results as well as the high vacuum performance. It also works stable during the whole progress of microwave commissioning.

## 1   Introduction

As one of the most widely used microwave components in many microwave systems, in most cases, the power divider is used to equally divide the input microwave power. In accelerators, power dividers are most needed when feeding a dual-feed accelerating structure or two single-feed accelerating structures next to each other.

Unlike the waveguide directional couplers [1-3] which are commonly used for on-time power monitoring and interlock protection. The most important performance of a waveguide power divider is to have two equal microwave power outputs on the aspects of both the magnitude and the phase, because either a dual-feed accelerating structure or two single-feed accelerating structures needs to be fed by two equal power to avoid the asymmetry of the electrical field inside their couplers [4].

A new accelerating structure with a dual-feed input coupler and a dual-output coupler has been developed for low beam emittance applications. As the most important part of the input waveguide for the accelerating structure, four kinds of waveguide power dividers with different structures working at 2.856 GHz are simulated and compared, finally the power divider with matching rod in the middle of its structure has been chosen because of having the best simulation performances. The prototype has got a qualified microwave measurement results and now is assembled together with the accelerating tube for microwave commissioning, the high power performance of the power divider is very stable.

## 2   Structure

Four kinds of waveguide power dividers all working at S band (2.856 GHz) are simulated in Computer Simulation Technology (CST), their structures are shown in Fig. 1 and are named as model 1 to 4 respectively. All the four models are simulated by using their vacuum part, and port 1 (input port) is on the top while port 2 and 3 (output ports) are on the left and right respectively. Port 1 in model 1 to 3 are matched by making part of the side

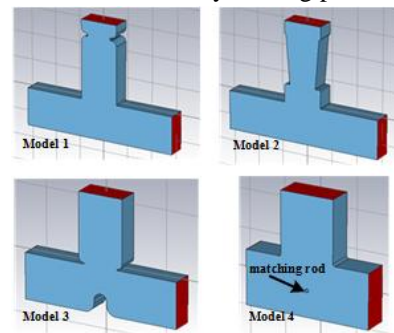

Fig.1 Structures (simulation models) of the four waveguide power dividers (3D)

---

*  Supported by National Natural Science Foundation of China (11505199)

1) E-mail: hexiang@ihep.ac.cn

wall narrower, while in model 4 it is matched by a matching rod in the middle of the structure.

## 3 Simulation and measurement result

The microwave power comes in from port 1 and are divided equally into port 2 and 3. In CST simulation, some microwave performances are compared and listed in Table 1 (VSWR refers to the Voltage Standing Wave Ratio), from which we can see that model 3 and 4 are better than model 1 and 2 with a wider VSWR bandwidth and a lower maximal electrical field.

Table 1. Comparison of simulation results of four power dividers working at 2.856 GHz

|  | Model 1 | Model 2 | Model 3 | Model 4 |
| --- | --- | --- | --- | --- |
| VSWR at 2.856 GHz | 1.039 | 1.043 | 1.058 | 1.018 |
| Bandwidth of VSWR less than 1.1 /MHz | 10.7 | 20.4 | 196.0 | 350.3 |
| Magnitude of S21 & S31 /dB | -3.062 (both) | -3.071 (both) | -3.019 (both) | -3.011 (both) |
| Phase difference of S21 & S31 /degree | 0 | 0 | 0 | 0 |
| Maximal electrical field of structure / V/m | 1245 | 1301 | 777 | 676 |

Finally, model 4 is chosen considering the simplicity of the structure and fabrication. The photo of the prototype is shown in Fig. 2 while the comparison between the simulation and measurement results are listed in Table 2.

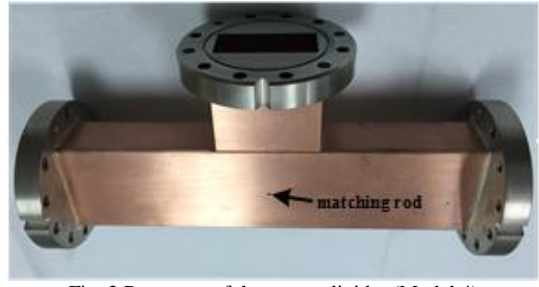

Fig. 2 Prototype of the power divider (Model 4)

Table 2. Comparison of simulation and measurement results of the power divider (Model 4)

|  | Simulation | Measurement |
| --- | --- | --- |
| VSWR at 2.856 GHz | 1.018 | 1.043 |
| Bandwidth of VSWR less than 1.1 /MHz | 350.3 | 70.8 |
| Magnitude of S21 & S31 /dB | -3.011 (both) | -3.13 (both) |
| Phase difference of S21 & S31 /degree | 0 | 0.5 |
| Maximal electrical field of structure / V/m | 676 | / |

From Table 2, we can see that the developed waveguide power divider has got a VSWR of 1.018/1.043 at central frequency (2.856 GHz), a bandwidth for VSWR less than 1.1 of 350.3/70.8 MHz, a magnitude for S21(S31) of -3.011/-3.13 dB as well as a phase difference of 0/0.5 degree for simulation/measurement results respectively. The fabrication error maybe the main reason for the worse measurement results compared to the simulation ones, however, the measurement results for this prototype is also qualified and acceptable.

The frequency response of VSWR and the simulation power flow are shown in Fig. 3 and 4 respectively, from which we can see that the power is divided equally into port 2 and 3.

The vacuum leakage test of the power divider has also been done, which shows a leakage better than $2*10^{-10}$ Torr·L/s. This vacuum performance is good enough for using and satisfies the designing goal.

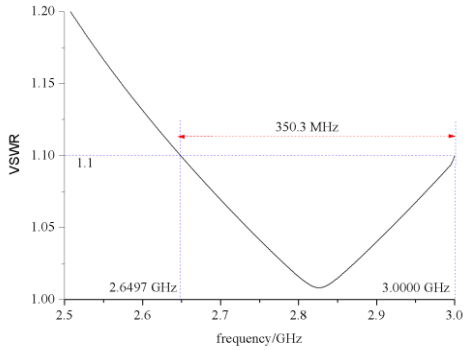

Fig. 3 Simulation frequency response of VSWR for the waveguide power divider

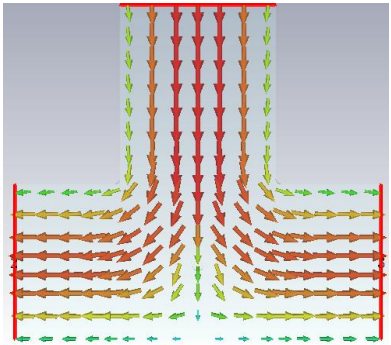

Fig. 4 Power flow for the waveguide power divider

## 4  Error analysis

For guiding the fabrication, the needed fabrication accuracy are calculated in CST and shown in Fig. 5. The calculating criterion is to guarantee that within this fabrication accuracy, the microwave performance of the waveguide power divider is still acceptable: VSWR at 2.856 GHz is less than 1.05, the bandwidth of VSWR less than 1.1 is more than 200 MHz, the magnitude of S21/S31 is better than -3.2 dB and the phase difference between S21 and S31 is less than 0.1 degree.

The total length ($L_1$) and width ($L_2$) of the waveguide power divider do not need to be very accurate, but should be longer than the given value. The other parameters all need an accuracy better than ±0.01 mm. The most important parameters are the diameter (D) and the distance to the waveguide edge (Offset) of the matching rod, for these two parameters affect the matching of the power divider directly.

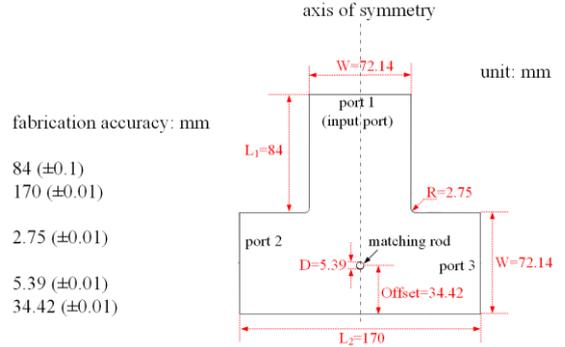

Fig. 5 Demand of fabrication accuracy of the waveguide power divider

## 5  Measure with the accelerating tube

In June of 2016, the waveguide power divider was assembled together with a dual-feed accelerating structure on the platform of high power test in the experimental hall of Beijing electron-positron collider II (BEPCII). The final goal for the microwave commissioning is to make the accelerating structure as well as the waveguides (including the power divider) working stable at 10 Hz (pulse repetition ratio) and 3 μs (pulse width), when the klystron gives a 40 MW (peak power, corresponding high voltage of the charging unit is 40 kV) output microwave power. Now the microwave commissioning has already approached 40 MW and the waveguide power divider is very stable during the whole progress.

Before the start of the microwave commissioning, we measured the VSWR of the accelerating structure from the input port of the waveguide power divider (Fig. 6) to make sure this system working normally at designed frequency. The measured VSWR at 2.856 GHz is 1.018 while the bandwidth of VSWR less than 1.2 is 4.7 MHz (2853.23~2857.93 MHz), which are very close to the measurement results of the cold test in accelerating structure measurement laboratory. The result indicates that the power divider is of good microwave performances from another angle.

## 6 Conclusion

Four kinds of waveguide power dividers with different structures are simulated in CST and the microwave performances are compared. Finally, the power divider with best simulation microwave performances are chosen for fabrication. The fabricated prototype has got qualified microwave and vacuum performances. The waveguide power divider is also qualified when measuring together with a dual-feed input accelerating structure, and being very stable during the whole progress of the microwave commissioning.

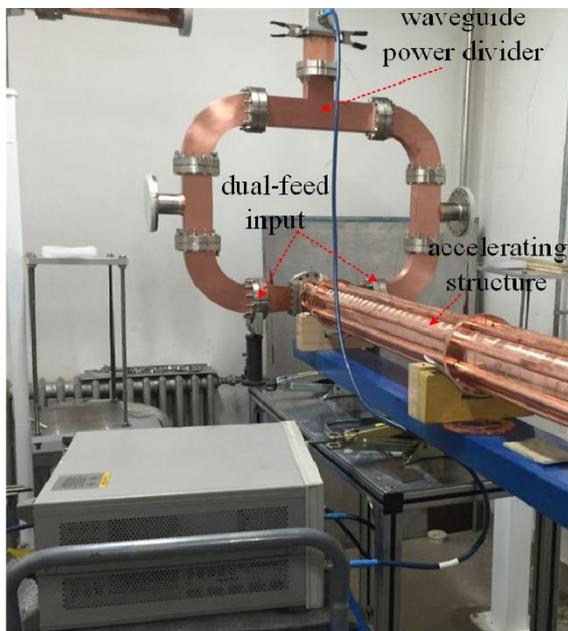

Fig. 6 The measurement of VSWR of the accelerating structure together with the power divider